\documentclass[useAMS,usenatbib]{mn2e}
\usepackage{graphicx,times}

\newcommand{\bm}[1]{\mbox{\boldmath $ #1 $}}

\newcommand{\qm}[1]{``#1''} 

\title[Bulges {\em versus} discs] {Bulges {\em versus} discs: the evolution 
of angular momentum in cosmological simulations of galaxy formation}
\author[J. Zavala, T. Okamoto, and Carlos S. Frenk]{Jesus
Zavala$^{1}$\thanks{Present address: Shanghai Astronomical Observatory, Nandan Road 80, Shanghai 200030, China; jzavala@shao.ac.cn}, Takashi Okamoto$^{2}$, Carlos S. Frenk$^{2}$\\ 
$^{1}$Instituto de Ciencias Nucleares, Universdidad Nacional
Aut\'onoma de M\'exico A.P. 70-543, M\'exico 04510  
D.F., M\'exico \\
$^{2}$Institute for Computational Cosmology, Department of Physics,
Durham University, South Road, Durham, DH1 3LE 
}
\begin{document}

\date{Accepted ---. Received ---; in original form ---}

\pagerange{\pageref{firstpage}--\pageref{lastpage}} \pubyear{2002}

\maketitle

\label{firstpage}

\begin{abstract}

We investigate the evolution of angular momentum in simulations of
galaxy formation in a cold dark matter universe. We analyse two model
galaxies generated in the N-body/hydrodynamic simulations of Okamoto et
al. Starting from identical initial conditions, but using different
assumptions for the baryonic physics, one of the simulations produced
a bulge-dominated galaxy and the other one a disc-dominated
galaxy. The main difference is the treatment of star formation and
feedback, both of which were designed to be more efficient in the
disc-dominated object. We find that the specific angular momentum of
the disc-dominated galaxy tracks the evolution of the angular momentum
of the dark matter halo very closely: the angular momentum grows as
predicted by linear theory until the epoch of maximum expansion and
remains constant thereafter. By contrast, the evolution of the angular
momentum of the bulge-dominated galaxy resembles that of the central,
most bound halo material: it also grows at first according to linear
theory, but 90\% of it is rapidly lost as pre-galactic fragments, into
which gas had cooled efficiently, merge, transferring their orbital angular
momentum to the outer halo by tidal effects. The disc-dominated galaxy
avoids this fate because the strong feedback reheats the gas which
accumulates in an extended hot reservoir and only begins to cool once
the merging activity has subsided. Our analysis lends strong support
to the classical theory of disc formation whereby tidally torqued gas
is accreted into the centre of the halo conserving its angular
momentum.

\end{abstract}


\section{Introduction}

In the standard theory of galaxy formation in the cold dark matter (CDM)
cosmogony, disc galaxies form when rotating gas cools inside a dark matter halo
and fragments into stars. Analytic and semi-analytic calculations of this
process start by assuming that tidal torques at early times impart the same
specific angular momentum to the gas and the halo and that the angular momentum
of the gas is conserved during the collapse
\citep{fall,momaw98,cole00,fa00}. It is unclear whether this assumption holds 
in gasdynamic simulations of galaxy formation.

The first attempts to simulate the formation of a spiral galaxy from
cold dark matter initial conditions generally failed, producing
objects with overly centrally concentrated distributions of gas and
stars. It became immediately apparent that the root cause of this
problem was a net transfer of angular momentum from the baryons to the
dark matter halo during the aggregation of the galaxy through mergers
\citep{NavBe,NavWhi,NavFreWhi}. This is known as the
\qm{angular momentum problem}.  It was suspected from the start that
its solution was likely to involve feedback processes that would
regulate the supply of gas to the galaxy.

More recent simulations within the CDM framework have produced more promising
disc galaxies. \citet{tc01} obtained a reasonably realistic disc at
$z=0.5$ (when their simulation stopped) by assuming
that gas cooling is strongly suppressed by feedback effects. 
\citet{SN02}, using different prescriptions for star formation
and feedback, found that a broad range of galaxy morphologies could be
produced. In related work, \citet{Abadi03} obtained a disc galaxy with
a realistic density profile but not enough angular momentum.
\citet{sgp03} found that considerably larger energy feedback 
rates at early times than previously employed were required to
ameliorate the angular momentum problem. They achieved this by
distinguishing between {\it early} and {\it late} star formation
modes, assuming a higher star formation efficiency and stronger
feedback in the early mode. They were able to generate a variety of
morphological types, including discs whose angular momentum, however,
was still a factor of two smaller than observed for real disc galaxies. 

\citet{gov04} produced a disc galaxy without requiring strong feedback and
ascribed the angular momentum problem to a lack of numerical
resolution. However, subsequently, \citet{gov06} were forced to invoke strong
feedback in order for their simulated discs to lie closer to the Tully-Fisher
relation. While resolution is clearly an important and complicated issue,
\cite{Oka1} argued that if the hot and cold gas components are decoupled from
each other, then the effects of resolution in the current generation of
simulations are reduced. \citet{rob04} adopted a multiphase model
for the star-forming interstellar medium (ISM) which stabilises gaseous discs
against the Toomre instability, and produced a galaxy with an exponential
surface brightness profile but insufficient angular momentum in the inner
parts. They concluded that the fragmentation of gaseous discs caused by the
Toomre instability is the main cause of the angular momentum loss in simulations
that usually assume an isothermal ISM at a temperature of $\sim 10^4$K.

\citet{Oka2} also adopted a multiphase model for the ISM and assumed a
top-heavy initial mass function (IMF) for stars formed in starbursts. The latter
assumption was motivated by the semi-analytic model of \citet{Baugh05} which
requires a large relative abundance of massive stars in order to match the
number counts of bright submillimeter galaxies. A top-heavy IMF is further
supported by the good agreement of this model with the metal abundances measured
in the intracluster medium \citep{nag05a} and in elliptical galaxies
\citep{nag05b}.  By varying the criteria for starbursts in their
simulations, \citet{Oka2} were able to produce galaxies with a variety
of morphological types, from ellipticals to spirals, starting from
exactly the same initial conditions. They showed that disc galaxies
can form in halos that have a relatively quiet merger history if the
early collapse of baryons is inhibited by strong feedback, in this
case generated by the evolution of a stellar population with a
top-heavy IMF.

In this paper, we analyse two of the galaxies simulated by
\citet{Oka2}, one which ended up as a bulge-dominated galaxy and
another that formed an extended stellar disc. We trace the time evolution of the
angular momentum of the dark matter and the gas, and we distinguish between the
evolution of the inner part of the dark halo and the halo as a whole. We show
explicitly that mergers transfer angular momentum from the inner parts of the
system to the outer halo. This process is one aspect of the redistribution of
angular momentum during virialization investigated in N-body simulations by
\citet{don06} and \citet{donghia07}. Thus, the spin of the {\em central} part of
the final system contains a fossil record of the merger history of the
object.

This paper~is organised as follows. The simulations are briefly described in
Section~2. The method we use to follow the angular momentum of the different
components is explained in Section~3 where we also present our results. We
conclude in section 4.

\section{The simulations}

Details of the simulations that we analyse here may be found in
\cite{Oka2}. The initial conditions were extracted from a cosmological
N-body simulation (of a cubical region of side $35.325 h^{-1}$~Mpc) and
correspond to a dark matter halo of present day virial mass $M_{\rm
vir} \simeq 1.2 \times 10^{12} h^{-1} M_{\odot}$ which has a quiet
merging history since $z \sim 1$. The simulations were carried out
using the N-body/SPH code {\sc gadget}-2 \citep{Springel05}. They
include a multiphase description of the star-forming gas and make use
of the `phase-decoupling' technique introduced by \citet{Oka1}
to suppress the spurious angular momentum transfer from cold disc
gas to ambient hot halo gas which would otherwise arise from the very
nature of SPH.

\begin{figure}
\centering
\includegraphics[height=7.5cm, width=7.5cm]{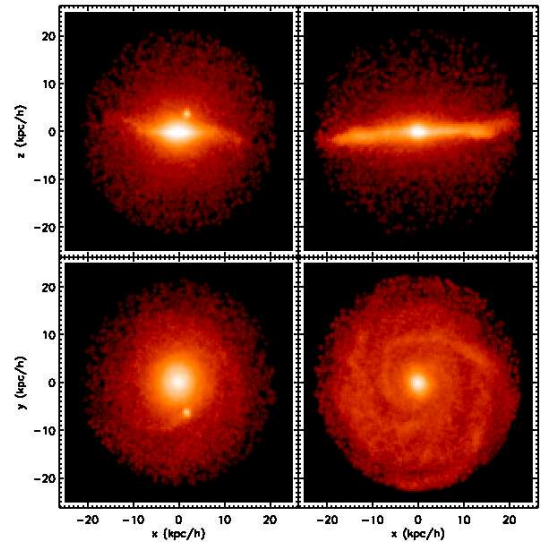}
\caption{Edge-on (top) and face-on (bottom) views of the baryonic component 
of the bulge-dominated (left) and disc-dominated (right)
galaxies. Stars and gas with
$\rho_g\ge7\times10^{-27}\textrm{gr~cm}^{-3}$ within $10\%$ of the
virial radius make up the baryonic component. The brightness indicates
the projected mass density.}
\label{galaxies}
\end{figure}
One of the two simulations that we analyse is the ``no-burst'' model
which has a standard prescription for star formation, tuned to
reproduce the local \cite{ken98} relation between star formation rate
and gas density.  This simulation yielded a galaxy with a large bulge
and a small disc, to which we will henceforth refer as the
``bulge-dominated'' galaxy (left panels in Fig.~\ref{galaxies}).  

The second simulation is the \qm{shock-burst} model in which the star formation
efficiency is higher than average in gas that has recently been
shock-heated. The stellar populations that form from this gas in a starburst are
assumed to have a top-heavy IMF. Thus, in this model there is additional
supernova heating available when galaxies merge. The outcome of this simulation
was a galaxy with a large disc and a small bulge -- the \qm{disc-dominated}
galaxy (right panels in Fig.~\ref{galaxies}). The (kinematically identified)
disc in this galaxy contains 48\% of the stellar mass and 84\% of the B-band
light, making this a good candidate for a late-type spiral. However, the surface
brightness distribution of the disc is considerably more extended than is
commonly observed for such galaxies \citep{Oka2}. Interestingly, the satellites
of the main galaxy in this model provide a good match to the properties of the
brightest Milky Way satellites \citep{lib07}.

\cite{Oka2} presented a third simulation in their paper, the
\qm{density-burst} model, which led also to a bulge-dominated object. 
We do not discuss our analysis of this object in this paper\ because
this galaxy lost most of its baryons in strong galactic winds and is
therefore not particularly suitable for studying the evolution of the
angular momentum of the baryonic component. Nevertheless, we have
checked that the results for this object are qualitatively similar to
those of the \qm{bulge-dominated} galaxy.

\begin{figure*} \centering \includegraphics[width=12cm]{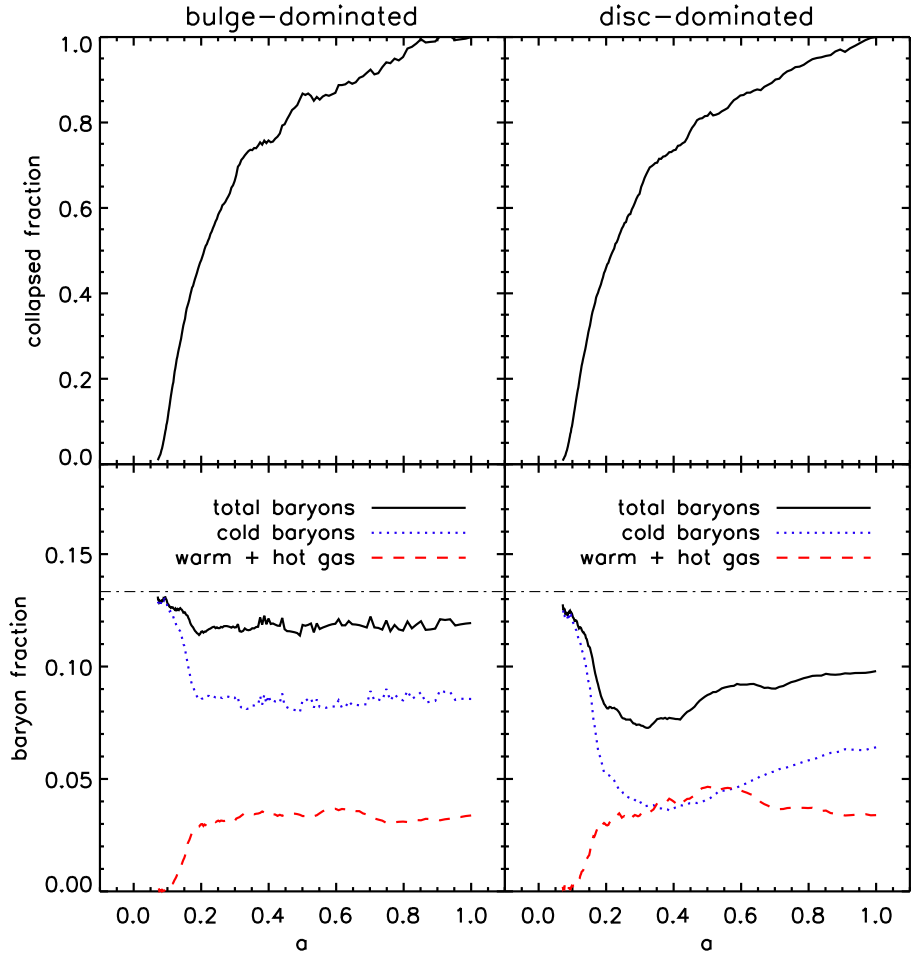}
\caption{Upper panels: the fraction of the final dark halo mass that
is contained in progenitor halos with particle number, $N_{\rm DM} \ge
10$, at redshift $z$ for the bulge-dominated (left) and the
disc-dominated (right) galaxies. Lower panels: baryon fractions in the
progenitor halos. The thick solid lines give the total mass fraction
of baryons in progenitor halos with $N_{\rm DM} \ge 10$. The dotted
line shows the contribution to this fraction from cold baryons defined
to be stars and dense gas with
$\rho_g\ge7\times10^{-27}\textrm{gr~cm}^{-3}$; the dashed line shows
the contribution of all other baryons to which we refer as warm/hot
gas. The thin line marks the cosmic baryon fraction, $\Omega_{\rm
b}/\Omega_0$. The left panel corresponds to the bulge-dominated galaxy
and the right panel to the disc-dominated galaxy.}  
\label{bfrac}
\end{figure*} 
In our analysis of angular momentum evolution, we
distinguish three (Lagrangian) components for each galaxy at
$z=0$. The centre of the system is taken to be the density peak of the
stellar distribution.  The first component is the dark matter halo
which we define as the dark matter mass contained within the virial
radius, $r_v$. We take $r_v$ to be the radius within which the mean
density is 100 times the critical value \citep{Eke}. For both
galaxies, the halo mass is $\sim 10^{12} h^{-1} M_{\odot}$ and the
virial radius is $\sim210~h^{-1}\textrm{kpc}$.  The second component
we consider is the {\em inner Lagrangian halo} which we define as the
10\% most bound dark matter particles at redshift $z = 0$. The third
component is the \qm{galaxy} which we define as the stars and dense
gas ($\rho \ge 7 \times 10^{-27} \textrm{g~cm}^{-3}$) contained within
$0.1 r_v$. In the next section, we will follow the evolution of these
three components.

Fig.~\ref{galaxies} shows edge-on and face-on views of the surface
density of the two galaxies.  The bulge-dominated galaxy (left) has a
centrally concentrated spheroidal bulge, a small disc and an extended,
nearly spherical stellar halo. The disc-dominated galaxy (right) has a
small central bulge, an extended disc and a spheroidal stellar halo.

In Fig.~\ref{bfrac} we highlight relevant aspects of the global
evolution of the Lagrangian region associated with the final dark
matter halo. We identify the progenitors of the final halo by
applying, at each redshift, the friends-of-friends grouping algorithm
of \citet{dav85} to the dark matter particles that end up in the halo
at $z=0$. We employ a linking length, $l_{\rm link} = b(z) \langle l
\rangle$, where $\langle l \rangle$ is the mean dark matter
interparticle separation and the linking parameter, $b(z)$, is scaled
from the canonical value of 0.2 for an Einstein-de-Sitter universe
according to 
\[b(z) = 0.2 \left( \frac{\Delta_{\rm v}(z)}{178}\right)^{-\frac{1}{3}},\] 
where $\Delta_{\rm v}(z)$ is the virial overdensity (relative to the mean) at
each redshift. As expected, the evolution of the dark matter component is very
similar for the two simulations. The halo builds up rapidly at first and, by
$z=4$, half of the final halo mass is already in place in progenitors of more
than 10 particles each. Large mergers are manifest as rapid increases in mass,
such as that visible at $z\sim 1.4$.

\begin{figure*}
\centering
\includegraphics[width=14cm]{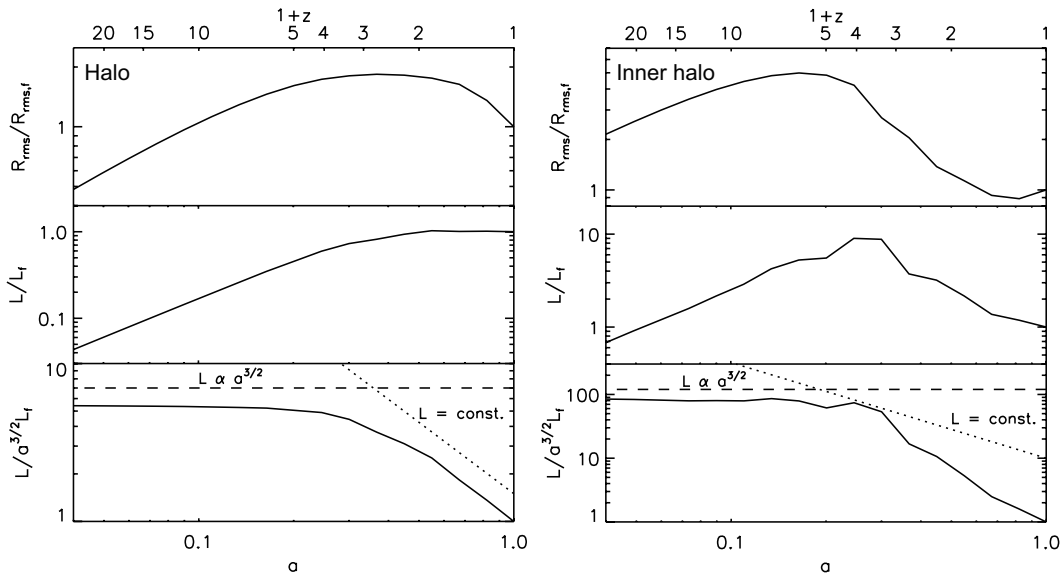}
\caption{Left: the evolution of the {\em rms} radius (top panel) and the
specific angular momentum (middle and bottom panels) of the dark matter
particles that lie within $r_v$ at $z = 0$. In the bottom panel the dashed line
indicates $L\propto a^{3/2}$ and the dotted line $L=\textrm{const}$. Right: as
the left but for the particles that make the inner dark matter halo as
defined in Section~2. All physical quantities are normalized to their values at
the present day. }
\label{dm_rv}
\end{figure*}
The evolution of the baryons in the collapsed regions is very different in the
two simulations. Initially, all the baryons are assumed to be cold but, as the
halo progenitors collapse, their associated baryons are shock-heated and begin
to cool radiatively. In the bulge-dominated case, the baryon fraction decreases
slightly at first but it soon settles into a quasi-steady state, with a cold
fraction (stars plus dense gas with
$\rho_g\ge7\times10^{-27}\textrm{gr~cm}^{-3}$) that is approximately 2.5 times
larger than the warm/hot fraction (the remaining baryons). The total baryon
fraction in the collapsed regions is in excellent agreement with the mean value
of $\sim 90\%$ of the cosmic mean found in non-radiative gas simulations of a
$\Lambda$CDM universe by
\citet{cra07}. By contrast, in the disc-dominated case, strong winds
eject almost half of the baryons from the progenitor halos by $z \sim
4$. Some of these, however, are recaptured after $z\sim 1.5$ and the
baryon fraction recovers to about $75\%$ of the cosmic mean by the
end. The warm/hot phase behaves similarly to the bulge-dominated case,
but the cold component is much lower, particularly during the crucial
period between $z=4$ and 1 when much of the merger activity is taking
place. The cold baryon fraction increases slightly towards the end,
reaching $\sim 0.5$ of the universal value at the end. In what
follows, we shall show how these differences in the history of the
baryon content influence the final morphology of
the galaxies in a profound way. 

\section{Results}

We now calculate the evolution of the physical (i.e. not the comoving)
angular momentum of the particles that make up each of the three
components defined in Section~2 at the present day: the total halo,
the inner halo and the galaxy. We identify the same particles at each
output time in the simulation, find the position and velocity of their
centre of mass and compute the physical angular momentum of each
system, $\bm{L}_T$, as

\begin{equation}\label{angular}
\bm{L}_T = \sum_i \bm{l}_i = \sum_i m_i (\bm{r}_i \times \bm{v}_i), 
\end{equation}
where $\bm{r}_i$ and $\bm{v}_i$ denote, respectively, the position and
velocity of particle $i$ relative to the centre of mass of the
system. The total specific angular momentum, $\bm{L}$, is given by
$\bm{L}=\bm{L}_T/M_T$, where $M_T$ is the total mass of the system.

\subsection{Dark matter component}

First, we focus in the dark matter component. The evolution of the
baryons has only a minor effect on the evolution of the halo and so
the results are very similar for the bulge- and disc-dominated
galaxies as already seen in Fig.~\ref{bfrac}. We show results only
for the latter case.

The top-left panel of Fig.~\ref{dm_rv} shows the evolution of the physical {\em
rms} radius of the dark matter halo as a function of the scale factor,
normalised to the present day {\em rms} radius. The evolution closely follows
the spherical collapse model: the system expands, reaches a maximum radius at $1
+ z \sim 3$ and begins to collapse at $1 + z \sim 2$ to form a virialized system
of size about half the radius at maximum expansion.  The middle-left panel shows
the evolution of the magnitude of the specific angular momentum, $L = |\bm{L}|$,
of all the dark matter particles that lie within $r_v$ at $z = 0$ as a function
of $a$, once again normalised to the present day value. During the rapidly
expanding phase, ($1 + z >4$), the system gains angular momentum in proportion
to $a^{3/2}$, as first calculated from tidal torque theory by \citet{White}.
During the collapsing phase, ($1 + z < 2$), the magnitude of the angular
momentum remains nearly constant. To show the dependence on the scale factor
more clearly, in the bottom-left panel of the figure we scale the specific
angular momentum by $a^{3/2}$. This plot clearly shows the two distinct
evolutionary phases, indicated by the dashed ($L\propto a^{3/2}$) and dotted
($L=\textrm{const}$) lines.  Our results agree well with previous analyzes of
N-body simulations \citep[e.g.][]{White, ct96}.

\begin{figure*}
\centering
\includegraphics[height=7cm, width=14.5cm]{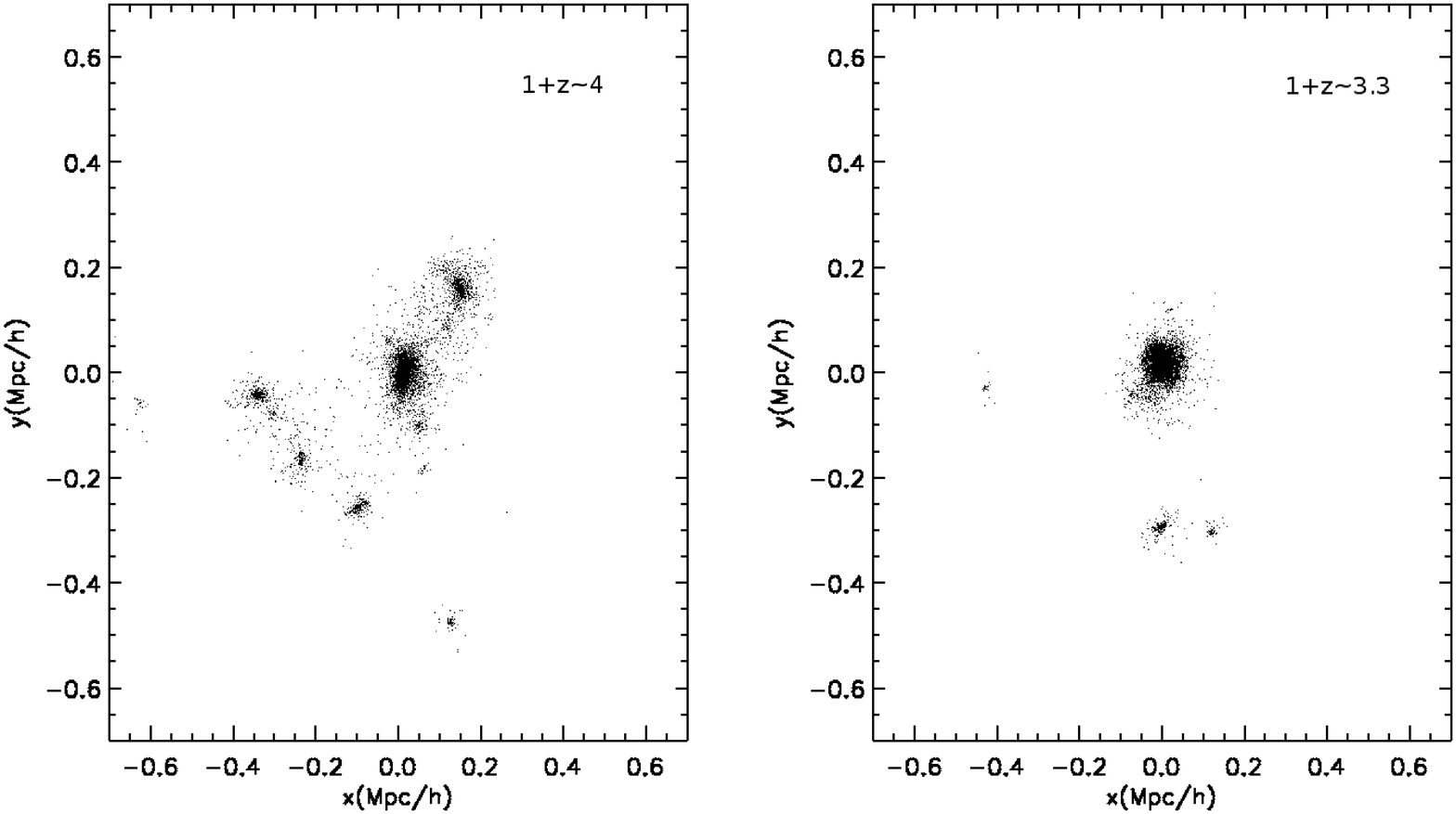}
\caption{Projected positions of the dark matter particles which will eventually
end up in the inner halo at $z=0$.  The left panel shows the
distribution at $1 + z\sim4$ and the right panel at $1+z\sim3.3$. }
\label{merger}
\end{figure*}

The right panels of Figure~\ref{dm_rv} show the same quantities as the left
panels but now for the particles that make up the inner dark matter halo
(i.e. the 10\% most bound mass) at $z=0$.  Since this subsystem has a
higher overdensity than the halo as a whole, it reaches maximum
expansion earlier and begins to collapse at $1 + z \sim 5$. The inner
halo shrinks more rapidly than the halo as a whole and ends up with a
radius that is only about $1/5$ of the size at maximum
expansion. At $1+z \sim 5$, the particles that will end up in the
inner halo are not contained in a single object, but in several
subclumps which subsequently merge. This is illustrated in
Fig.~\ref{merger} which shows the spatial distribution of the inner
halo particles at $1 + z \simeq 4$ and $3.3$. Many of the fragments
merge between these two epochs and it is these mergers that determine
the evolution of the {\em rms} size of the system during this period.

The middle and bottom panels of Fig.~\ref{dm_rv} show that the evolution of the
specific angular momentum of the inner halo is very different from that of the
halo as a whole. Initially, during the expansion phase, the system gains angular
momentum in much the same way as the halo as a whole, roughly following the
$L\propto a^{3/2}$ scaling. However, after maximum expansion, ($1 + z < 4$), the
inner halo rapidly looses most of its angular momentum. This behaviour is due to
the intense merging activity illustrated in Fig.~\ref{merger}. As fragments are
drawn towards the centre by dynamical friction, the asymmetric distribution of
dark matter produced by gravitational tides exerts a torque which transfers the
orbital angular momentum of the fragments to the outer halo. Each merger event
is accompanied by a decline in the angular momentum and in this way 90\% of the
angular momentum is drained from the inner halo.  During this phase, the angular
momentum of this system declines roughly as $a^{-3}$. This behaviour was first
noted in the early cold dark matter simulations of \citet{Frenk85}. The
redistribution of angular momentum during virialization has been studied in
detail by
\citet{donghia07}.

\subsection{Bulge-dominated galaxy}

We now perform the same analysis on the baryonic component of the
bulge-dominated galaxy illustrated in the left panels of
Fig.~\ref{galaxies}.  The total stellar mass of this system is
$6.75\times10^{10}h^{-1}M_{\odot}$ and the total gas mass is
$7.5\times10^{9}h^{-1}M_{\odot}$.

\begin{figure*}
\centering
\includegraphics[width=15cm]{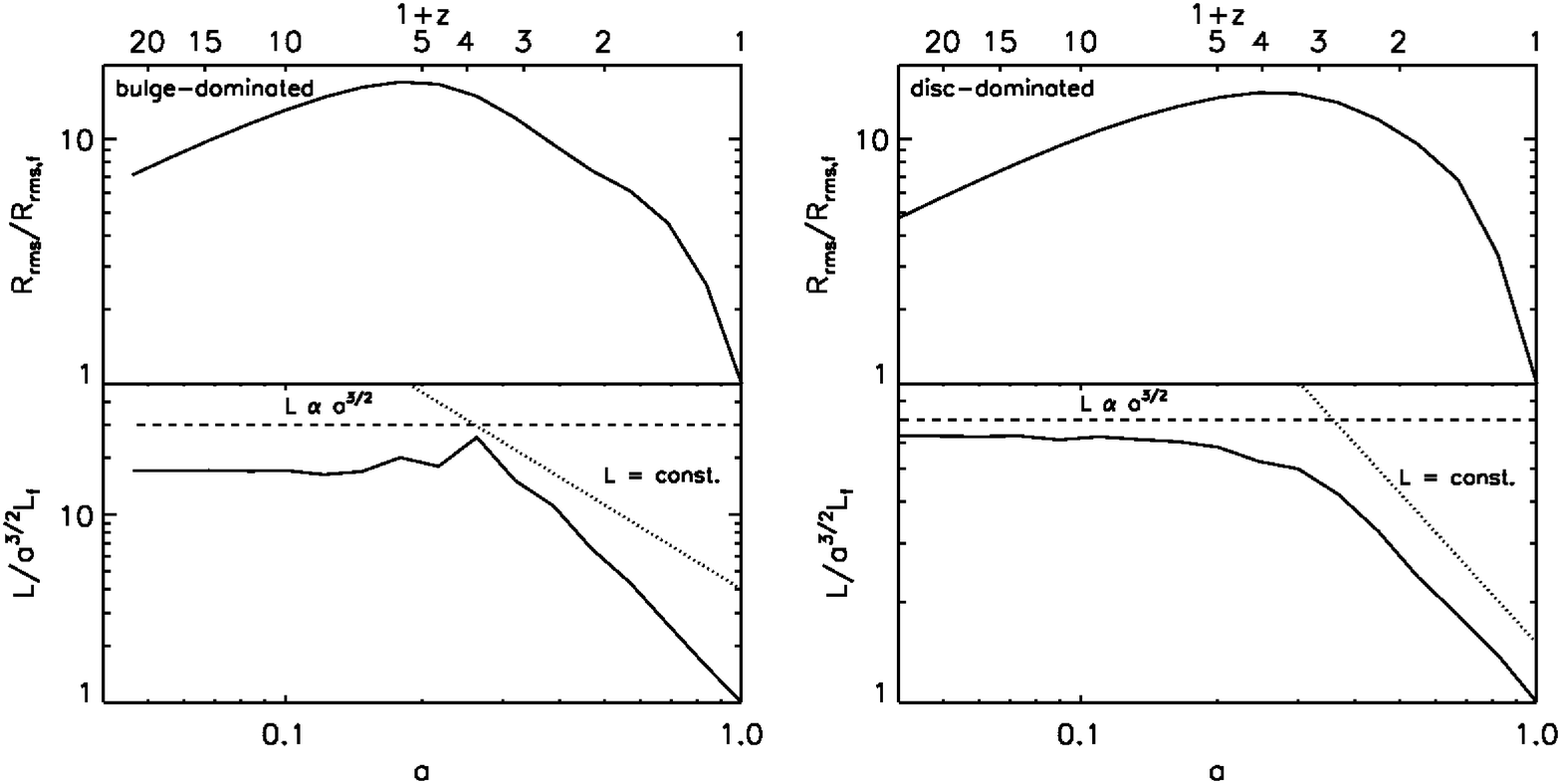}
\caption{Evolution of the {\em rms} radius (upper panel) and specific
angular momentum (lower panel) with scale factor for the
baryonic component. The left panels correspond to the bulge-dominated galaxy
and the right panels to the disc-dominated galaxy. All physical
quantities are normalized to their value at the present day.}
\label{bulge-dominated_ang}
\end{figure*}

The evolution of the {\em rms} radius of this galaxy, plotted in the
upper-left panel of Fig.~\ref{bulge-dominated_ang}, is remarkably
similar to that of the inner dark matter halo. The system expands
until $1 + z \sim 5$ and then collapses. The evolution of the magnitude of the
specific angular momentum (lower-left panel) is also very similar to that
of the inner dark matter halo.  During the expanding phase, the system
gains angular momentum in proportion to $a^{3/2}$, as expected from
linear theory. After $1 + z \simeq 4$, the system rapidly loses its
angular momentum, just as the inner dark matter halo did, transferring
it to the outer halo. The decline in the specific angular momentum of
the galaxy during the collapsing phase is less abrupt than for the
inner halo, scaling roughly a $L \propto a^{-1}$.

The similarity in the behaviour of the bulge-dominated galaxy and the
inner halo arises because even before the collapse phase, the cold
baryonic material is distributed in a similar way to the mass of the
inner halo, as may be seen in Fig.~\ref{bfrac}. Radiative cooling of
gas is very efficient in dense subclumps at early times and, in this
simulation, much of the gas cools and turns into stars inside the
fragments that will later merge to make the inner halo. Very little
gas is left over for later accretion and, as shown in
Fig.~\ref{bfrac}, the cold gas fraction hardly changes between $z=4$
and the present. Just as in the dark matter only case, the fragments,
now containing a mixture of dark matter and cold baryons (stars and
dense gas), merge together transferring their orbital angular momentum
to the outer halo. With most of the final cold baryons already in
place within the fragments at early times, the end result of the
collapse and merger phase is a bulge-dominated object.

\subsection{Disc-dominated galaxy}

We now analyse the disc-dominated galaxy illustrated in the right-hand panels of
Fig.~\ref{galaxies}. This galaxy has $4.17\times10^{10} h^{-1} M_{\odot}$ in
stars and $1.53\times10^{10} h^{-1} M_{\odot}$ in cold dense gas. The results are
shown in the right hand panels of Fig.~\ref{bulge-dominated_ang}.

The top-right panel of Fig.~\ref{bulge-dominated_ang} shows that the baryonic
material that will end up in the galaxy expands until $1 + z \simeq 4$
and then begins to collapse. This happens slightly later than for
the bulge-dominated galaxy. The lower-right panel shows that, as for all
other objects and components, the evolution of the angular momentum
during the expanding phase follows linear theory closely, with $L
\propto a^{3/2}$. However, unlike the bulge-dominated galaxy, the
angular momentum of the system remains nearly constant during the collapsing
phase ($1+z < 3$), indicating that the accreting gas conserves its angular
momentum. This behaviour is reminiscent of that of the halo as a whole,
illustrated in Fig.~\ref{dm_rv}. Indeed, the lower-right panel of
Fig.~\ref{bulge-dominated_ang} is very similar to the bottom-left panel of
Fig~\ref{dm_rv}.

The correspondence between the evolution of the angular momentum of
the baryonic material of the disc-dominated galaxy and of the material
that makes up the halo follows from the similarity of their spatial
distributions at the time of maximum expansion. As discussed in
Section~2, mergers of subclumps in this simulation induce large-scale
shocks in the gas which result in an increased star formation
efficiency. This, in turn, produces strong feedback which is further
promoted by the assumed top-heavy starburst IMF, keeping the gas out
of the merging fragments. Since the merger rate is higher at high
redshift, this strong feedback suppresses the early collapse of gas
into small proto-galaxies. The process is apparent in
Fig.~\ref{bfrac}: the wind-driven ejection of gas at $z\sim 4$
drastically reduces the cold gas fraction in progenitor halos which
remains much smaller than in the bulge dominated galaxy. Thus, most of
the baryons that will make up the final galaxy become decoupled from
the sub-clumps early on and accumulate in a hot gas reservoir in the
main halo. This hot gas has similar specific angular momentum than the
halo and, as it looses its pressure support through radiative cooling,
it accretes onto the galaxy conserving its angular momentum. The key
to the formation of disc galaxies is hence the suppression of the
early collapse of baryons into merging fragments, as anticipated by
many authors \citep{NavSte, Oka2, Sommer, tc01, Weil}.

\begin{figure}
\centering
\includegraphics[width=6cm]{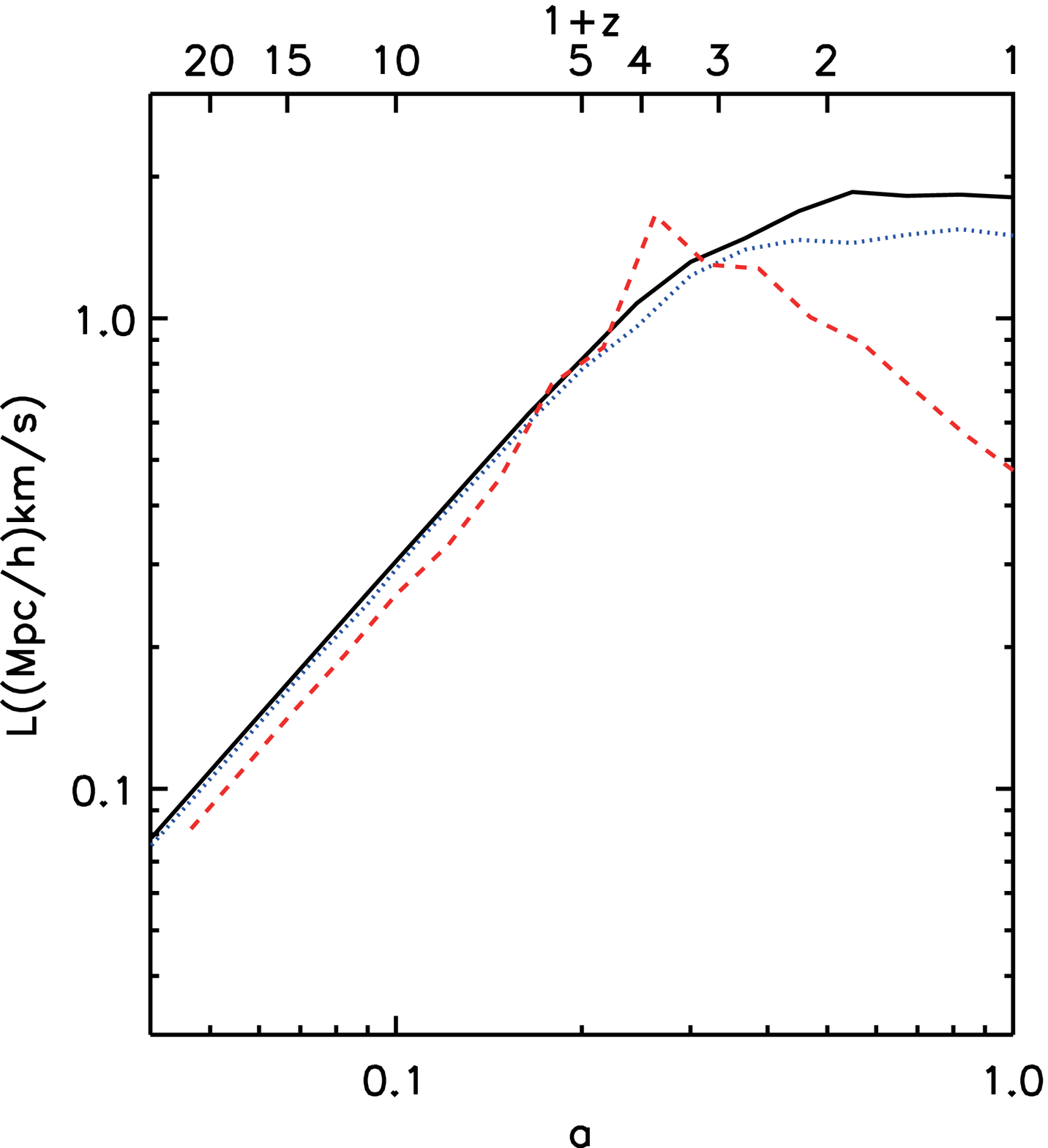}
\caption{Specific angular momentum of the dark matter component
(solid line) and of the baryonic component of the bulge-dominated
galaxy (dashed line) and of the disc-dominated galaxy (dotted line).}
\label{dm_bar}
\end{figure}

Finally, in Fig.~\ref{dm_bar} we show the evolution of the
unnormalised specific angular momentum of the whole halo (solid line)
and of the baryonic components of both galaxies, the bulge-dominated
(dashed line) and the disc-dominated (dotted line) objects.  This
figure indicates that it is not only the shape of the time dependence
of the angular momentum that is similar between the halo and the
baryons of the disc-dominated galaxy, but, remarkably, also the value
of the specific angular momentum of the two components.  Thus, our
simulation provides strong support for the classical theory of disc
formation whereby tidally torqued gas is accreted into the centre of
the halo conserving its angular momentum \citep{fall,momaw98}. This is
the assumption made in semi-analytic models of galaxy formation 
\citep[e.g.][]{cole00}. By contrast, the bulge-dominated galaxy ends up
with only about 25\% of the specific angular momentum of its halo.

\section{Discussion and conclusions}

The inability to obtain realistic discs by the present day in simulations from
cold dark matter initial conditions has been ascribed to various possible
causes: (i) a lack of numerical resolution \citep{kau07, gov04, gov06}, 
(ii) an artificial transfer of angular momentum from a cold disc to a hot 
surrounding atmosphere \citep{Oka1}, (iii) instabilities in a heavy disc 
which, by causing fragments to form, end up transporting angular momentum 
inwards \citep{rob04}, and (iv) the transfer of angular from
merging substructures to the outer halo \citep{NavBe,NavWhi,NavFreWhi}. 
The first three of these four considerations are unimportant in our 
simulations. \cite{Oka1} carried out a convergence study which suggests that, 
so long as the cold and hot gas phases are dynamically decoupled, resolution 
effects are not important in calculations the size of those we analyse here. 
Such decoupling also suppresses process (ii). 
The adoption of an effective multiphase equation-of-state stabilizes a
disc against gravitational instability thus alleviating (iii). Process (iv) is
inevitable and is the dominant form of angular momentum transport (at least for
the dark matter) in our simulations.

The series of N-body/SPH simulations of galaxy formation in a CDM universe
carried out by \citet{Oka2} led to the conclusion that galaxies with very different
morphologies can form in the same dark matter halo depending on the details of
the assumed star formation and feedback prescriptions. In this paper, we have
investigated the evolution of the specific angular momentum as a possible
explanation for why relatively small differences in the assumed baryonic
processes can result in such different outcomes. We have analyzed the
bulge-dominated and the disc-dominated galaxies simulated (from identical
initial conditions) by \citet{Oka2}. We trace back all the dark matter particles
that end up within the virial radius of the halo ($r_v$), as well as those that
end up in the central parts of the halo by virtue of being in the top 10
percentile of the binding energy distribution. We also trace back the baryonic
particles that end up as a galaxy, i.e. as cold baryons (star particles and
dense gas) within 10\% of $r_v$ at $z = 0$.

The angular momentum of the dark matter halo evolves almost
identically in the two simulations. The specific angular momentum of
the halo as a whole grows initially as predicted by tidal torque
theory in the linear regime, in proportion to $a^{3/2}$
\citep{White, ct96}. This phase lasts until maximum expansion is
reached. Thereafter, the angular momentum of the dark matter remains
constant, as the system collapses to a virialized configuration. By
contrast, the evolution of the most bound 10\% of the dark matter is
strikingly different. In the linear phase this material behaves just
as the halo as a whole but, after maximum expansion, its specific
angular momentum rapidly declines to $\sim 10\%$ of its maximum
value. The decline occurs in episodes associated with the mergers
that build up the central halo. Most of the angular momentum of these
fragments is invested in their orbits and is transferred to the outer
halo by tidal forces as the fragments sink by dynamical friction.

The baryonic component evolves very differently in the two
simulations. In essence, in the disc-dominated galaxy, the specific
angular momentum of the baryons tracks that of the halo as a whole,
while in the bulge-dominated galaxy, it tracks the evolution of the
central halo material. The reason for the difference can be traced
back to the different strengths of feedback in the two galaxies at
early times. In the bulge-dominated galaxy, feedback is weak and most
of the baryons rapidly cool and condense into stars within
sub-galactic fragments. As these halo fragments merge and give up
their angular momentum to the outer halo, so do the baryon clumps
within them. The result is a slowly-rotating central bulge. A small
amount of residual gas rains in later on forming a small disc. In the
disc-dominated galaxy, by contrast, feedback is strong at early times,
the gas is reheated before it can make a substantial mass of stars
and, instead, accumulates in an extended hot reservoir which acquires
a specific angular momentum similar to that of the dark halo. By the
time this gas is able to cool, much of the merger activity has
subsided and so the gas is able to dissipate and collapse into an
centrifugally supported disc configuration conserving its angular
momentum. Not only does the angular momentum of the baryonic material
track the evolution of the angular momentum of the halo, but the
actual values of the specific angular momentum of both components are
very similar throughout the entire history of the galaxy. By contrast,
the specific angular momentum of the bulge-dominated galaxy is only
about 25\% of the halo value.

Our results confirm that the key to the formation of disc galaxies in the cold
dark matter cosmology is the suppression of the early collapse of baryons into
small, dense halos. By contrasting the different evolutionary histories of the
angular momentum in bulge- and disc-dominated galaxies, our analysis explicitly
reveals how regulating the rate at which gas is supplied determines the final
morphology of the galaxy. A disc galaxy results when, as seen in
Fig.~\ref{bfrac}, a large fraction of the baryon content of the halo is ejected
by winds at early times. The ejected baryons (some of which are recaptured
later) are thus decoupled from the subhalos that are destined to merge and join
a hot gas reservoir which is tidally torqued in the same way as the host halo
and thus acquires the same net specific angular momentum. Our results provide
strong support for the classical theory of disc formation by \citet{fall} and
\citet{momaw98} in which tidal torques impart the same specific
angular momentum to the dark matter and the gas and angular momentum is
conserved during disc formation.

\section*{Acknowledgments}

This work was carried out during a research visit of JZ to the ICC at
Durham supported by the EU's ALFA programme through the Latin American
European Network for Astrophysics and Cosmology. JZ acknowledges
support from CONACyT and DGEP-UNAM scholarships. We thank Simon White
for insightful comments. TO and CSF acknowledge support from
PPARC. CSF acknowledges receipt of the Royal Society Wolfson Research
Merit Award. The simulations were carried out at the Cosmology Machine
at the ICC.

\bsp

\label{lastpage}


\begin{thebibliography}{99}
\bibitem[\protect\citeauthoryear{Abadi et al.}{2003}]{Abadi03} 
Abadi M.~G., Navarro J.~F., Steinmetz M., Eke V.~R., 2003, ApJ, 591, 499 
\bibitem[\protect\citeauthoryear{Baugh et al.} {2005}]{Baugh05} Baugh,
C.~M., Lacey, C.~G., Frenk, C.~S., Granato, G.~L., Silva, L., Bressan,
A., Benson, A.~J., \& Cole, S.\ 2005, MNRAS, 356, 1191 
\bibitem[\protect\citeauthoryear{Catelan \& 
  Theuns}{1996}]{ct96} Catelan P., Theuns T., 1996, MNRAS, 
  282, 436 
\bibitem[Cole et al. (2000)]{cole00} Cole, S., Lacey, C.~G., 
Baugh, C.~M., \& Frenk, C.~S.\ 2000, MNRAS, 319, 168 

\bibitem[\protect\citeauthoryear{Crain et al.}{2007}]{cra07} 
Crain R.~A., Eke V.~R., Frenk C.~S., Jenkins A., McCarthy I.~G., Navarro 
J.~F., Pearce F.~R., 2007, MNRAS, 377, 41 
\bibitem[\protect\citeauthoryear{{Davis}, {Efstathiou}, {Frenk} \&
    {White}}{{Davis} et~al.}{1985}]{dav85}
{Davis} M.,  {Efstathiou} G.,  {Frenk} C.~S., {White} S.~D.~M.,  1985, ApJ,
    292, 371

\bibitem[\protect\citeauthoryear{D'Onghia et 
      al.}{2006}]{don06} D'Onghia E., Burkert A., Murante G., 
      Khochfar S., 2006, MNRAS, 372, 1525 

\bibitem[\protect\citeauthoryear{D'Onghia \& 
Navarro}{2007}]{donghia07} D'Onghia E., Navarro J.~F., 2007, MNRAS, 380, L58

\bibitem[\protect\citeauthoryear{Eke, Cole, \& 
    Frenk}{1996}]{Eke} Eke V.~R., Cole, S.~M., Frenk C.~S., 
    1996, MNRAS, 282, 263

\bibitem[\protect\citeauthoryear{Fall \& Efstathiou}{1980}]{fall} Fall S.~M.,
Efstathiou G., 1980, MNRAS, 193, 189  

\bibitem[Frenk et al.(1985)]{Frenk85} Frenk, C.~S., White, 
S.~D.~M., Efstathiou, G., \& Davis, M.\ 1985, Nature, 317, 595 

\bibitem[\protect\citeauthoryear{Firmani \& 
  Avila-Reese}{2000}]{fa00} Firmani C., Avila-Reese V., 2000, 
  MNRAS, 315, 457 

\bibitem[\protect\citeauthoryear{Governato et 
    al.}{2004}]{gov04} Governato F., et al., 2004, ApJ, 607, 688 

\bibitem[\protect\citeauthoryear{Governato et 
      al.}{2007}]{gov06} Governato F., Willman B., Mayer L., Brooks 
      A., Stinson G., Valenzuela O., Wadsley J., Quinn T., 2007, MNRAS, 374, 1479 
      
\bibitem[\protect\citeauthoryear{Kaufmann et 
  al.}{2007}]{kau07} Kaufmann T., Mayer L., Wadsley J., Stadel 
        J., Moore B., 2007, MNRAS, 375, 53 

\bibitem[\protect\citeauthoryear{Kennicutt}{1998}]{ken98} 
  Kennicutt R.~C., Jr., 1998, ApJ, 498, 541 

\bibitem[\protect\citeauthoryear{Libeskind et 
  al.}{2007}]{lib07} Libeskind N.~I., Cole S., Frenk C.~S., 
  Okamoto T., Jenkins A., 2007, MNRAS, 374, 16

\bibitem[\protect\citeauthoryear{Mo et al.}{1998}]{momaw98} Mo, H.~J., Mao, S.,
\&  White, S.~D.~M.\ 1998, MNRAS, 295, 319   

\bibitem[\protect\citeauthoryear{Nagashima et 
    al.}{2005a}]{nag05a} Nagashima M., Lacey C.~G., Baugh C.~M., 
    Frenk C.~S., Cole S., 2005a, MNRAS, 358, 1247 

\bibitem[\protect\citeauthoryear{Nagashima et 
  al.}{2005b}]{nag05b} Nagashima M., Lacey C.~G., Okamoto T., 
  Baugh C.~M., Frenk C.~S., Cole S., 2005b, MNRAS, 363, L31 

\bibitem[\protect\citeauthoryear{Navarro \& 
        Benz}{1991}]{NavBe} Navarro J.~F., Benz W., 1991, ApJ, 380, 
        320 
\bibitem[\protect\citeauthoryear{Navarro, Frenk, \& 
    White}{1995}]{NavFreWhi} Navarro J.~F., Frenk C.~S., White 
    S.~D.~M., 1995, MNRAS, 275, 56 
\bibitem[\protect\citeauthoryear{Navarro \& 
      White}{1994}]{NavWhi} Navarro J.~F., White S.~D.~M., 1994, 
      MNRAS, 267, 401 
\bibitem[\protect\citeauthoryear{Navarro \& 
        Steinmetz}{2000}]{NavSte} Navarro J.~F., Steinmetz M., 2000, 
        ApJ, 538, 477 
\bibitem[\protect\citeauthoryear{Okamoto et 
  al.}{2005}]{Oka2} Okamoto T., Eke V.~R., Frenk C.~S., 
  Jenkins A., 2005, MNRAS, 363, 1299 
\bibitem[\protect\citeauthoryear{Okamoto et 
    al.}{2003}]{Oka1} Okamoto T., Jenkins A., Eke V.~R., Quilis 
    V., Frenk C.~S., 2003, MNRAS, 345, 429 
\bibitem[\protect\citeauthoryear{Robertson et 
  al.}{2004}]{rob04} Robertson B., Yoshida N., Springel V., 
Hernquist L., 2004, ApJ, 606, 32 
\bibitem[\protect\citeauthoryear{Sommer-Larsen \& 
  Dolgov}{2001}]{Sommer} Sommer-Larsen J., Dolgov A., 2001, ApJ, 
  551, 608 
\bibitem[\protect\citeauthoryear{Sommer-Larsen, G{\"o}tz, \& 
        Portinari}{2003}]{sgp03} Sommer-Larsen J., G{\"o}tz M., 
        Portinari L., 2003, ApJ, 596, 47 
\bibitem[Springel(2005)]{Springel05} Springel, V.\ 2005, MNRAS, 
364, 1105 
\bibitem[\protect\citeauthoryear{Steinmetz \& 
Navarro}{2002}]{SN02} Steinmetz M., Navarro J.~F., 2002, 
NewA, 7, 155 
\bibitem[\protect\citeauthoryear{Thacker \& 
  Couchman}{2001}]{tc01} Thacker R.~J., Couchman H.~M.~P., 
  2001, ApJ, 555, L17 
\bibitem[\protect\citeauthoryear{Weil, Eke, \& 
  Efstathiou}{1998}]{Weil} Weil M.~L., Eke V.~R., Efstathiou 
  G., 1998, MNRAS, 300, 773 
\bibitem[\protect\citeauthoryear{White}{1984}]{White} White 
S.~D.~M., 1984, ApJ, 286, 38 
\end{thebibliography}
\end{document}